# Lepton Flavour Violating $\Lambda_b$ decays in non-universal $Z'$ model

S. Biswas[1], P. Nayek, P. Maji and S. Sahoo[2]

Department of Physics, National Institute of Technology Durgapur
Durgapur-713209, West Bengal, India
[1]E-mail: getswagata92@gmail.com, [2]E-mail: sukadevsahoo@yahoo.com

**Abstract**

Motivated by the recent LHCb results of lepton flavour violation on $b \to s$ and $b \to c$ transitions we study the lepton flavour violating (LFV) baryonic decays $\Lambda_b \to \Lambda l_i^+ l_j^-$ in non-universal $Z'$ model. We discuss the two-fold decay distribution of $\Lambda_b \to \Lambda l_i^+ l_j^-$ decays in terms of transversity amplitudes. From this distribution we study the differential branching ratio and lepton side forward-backward asymmetry in new physics (NP). The predicted values of the observables are very interesting and that might emboss the footprints of NP more aesthetically.

## I.     Introduction

Recent results by the LHCb, BaBar and Belle collaborations in neutral and charged current transitions of the mesons containing $b$ hadrons arouse curiosity about lepton flavour universality (LFU) violation. These LFV decay processes are extremely suppressed in the Standard Model (SM) because the expected levels at the SM lie far below current experimental sensitivities. In particular the branching fractions of $B^0 \to \tau^\pm \mu^\mp$ and $B_s \to \tau^\pm \mu^\mp$ decays are obtained in SM of order $10^{-54}$ [1] whereas experimentally they are constrained at the order of $10^{-5}$ by BaBar and LHCb with 90% and 95% confidence level respectively [2, 3]. Actually in the SM the generation lepton number of electroweak interactions is conserved but the observation of neutrino oscillation indicates the family lepton number violation. In the charged current interaction of the W boson, the mismatch between weak and mass eigenstates of neutrinos causes mixing between different generations of leptons [4]. Due to LFV, the flavour changing neutral current (FCNC) transitions in lepton sector should analogous to the principle occur for quark sector. In these FCNCs the mixing in the charged current interaction with the left-handed W boson and tiny neutrinos are extremely small because they are suppressed by powers of $m_\nu^2/m_W^2$. Other observables that are used to test LFU with the FCNC $b \to sll$ are: $R_{K^{(*)}} = \frac{Br(B \to K^{(*)} \mu^+ \mu^-)}{Br(B \to K^{(*)} e^+ e^-)}$. The most recent result established by LHCb is $(R_K)_{new} = 0.846^{+0.060+0.016}_{-0.054-0.014}, 1 \leq q^2 \leq 6$ GeV$^2$ [5] where $q^2$ is dilepton mass squared. This result is lower than the SM prediction $(R_K)_{SM} = 1.00 \pm 0.01$ [6] by $2.5\sigma$ discrepancy. $R_{K^*}$ is also experimented recently at LHCb and the value set as [7]:

$$R_{K^*} = \begin{cases} 0.66^{0.11}_{-0.17} \pm 0.03, 0.045 \leq q^2 \leq 1.1 \\ 0.69^{0.11}_{-0.07} \pm 0.05, 1.1 \leq q^2 \leq 6.0 \end{cases}$$

Here $q^2$ is in unit GeV$^2$. These results are deviated from SM predictions $R_{K^*}^{SM} = 0.906 \pm 0.028$ and $R_{K^*}^{SM} = 1.00 \pm 0.01$ by $2.3\sigma$ and $2.5\sigma$ discrepancies respectively. Belle has also set their values of $R_K$ and $R_{K^*}$ which are closer to the SM [8] value but with high uncertainties. Not only the FCNC transitions $b \to sll$ but also the charged current transition $b \to cl\upsilon$ hints the LFU violation with the observable $R_D$ and $R_{D^*}$. Belle [9-11], BaBar [12] and LHCb [13] has measured $R_{D^*}$. The new measurement of Belle [14] set the values as:

$$R_D = 0.307 \pm 0.37 \pm 0.016\,,$$
$$R_{D^*} = 0.283 \pm 0.018 \pm 0.14\,.$$

These results are greater than the SM predictions given in ref. [15] and ref. [16] by $2.3\sigma$ and $3.4\sigma$ deviations respectively.

However there are several theoretical models proposed to explain possible experimental tensions within current sensitivity discussed above. Therefore it can be said that the models that generate LFU violation also can generate LFV processes which are prohibited strictly at the SM. Various lepton flavour violating decays, such as $\tau \to 3\mu, \mu \to 3e, l \to l'M$ (where $l, l'$ are different leptons and $M$ is meson) and radiative decays $\mu \to e\gamma$ etc are studied in different NP models though there are no direct experimental evidence of these decays but their experimental bounds exist. These decays previously explained with the effect of FCNC mediated $Z$ boson [4, 17], in non-universal $Z'$ model [18-20], in leptoquark model [21-24], in MSSM [25-27] and other NP models [28] and also in model independent way [29]. In ref. [30] various NP operators of LFV decays are analysed with optimal observable technique. The LFV decays in B meson and charged lepton sector are extensively investigated; therefore it will be very interesting to observe lepton flavour violation in baryonic decays. In the present paper we have studied LFV baryonic decay $\Lambda_b \to \Lambda l_i^+ l_j^-$ induced by the quark level transition $b \to sl_i^+ l_j^-$ where $l_i^+$ and $l_j^-$ are charged leptons of different flavours. The analogue part $\Lambda_b \to \Lambda ll$ is observed in LHCb [31, 32] but all we know that currently there is no experimental data on $\Lambda_b \to \Lambda l_i^+ l_j^-$. Contrasted with $\Lambda_b \to \Lambda ll$, the favourable fact about $\Lambda_b \to \Lambda l_i^+ l_j^-$ decay is that it does not suffer from long distance QCD and charmonium resonance effects. Previously we have studied $\Lambda_b \to \Lambda ll$ decays in non-universal $Z'$ model [33] and here we study the differential branching fractions and forward-backward asymmetries of $\Lambda_b \to \Lambda l_i^+ l_j^-$ decays in this NP model of extended gauge group.

To understand physics beyond the SM, non-universal $Z'$ model is one of the most important and appreciated theoretically models [34-38]. In this model, the NP is allowed to contribute at tree level by $Z'$-mediated flavour changing $b \to q(q = s, d)$ decays where $Z'$ boson couples to the flavour-changing part $\bar{q}b$ as well as to the leptonic part $l_i^+ l_j^-$. Traditionally in different grand unified theories (GUTs) the mass of the $Z'$ boson is taken as arbitrary as it is not discovered till now. That is why various experiments and detectors have constrained the $Z'$ mass restricting its upper and lower limit. Different accelerators set the model-dependent lower bound about 500 GeV [39-41]. The range of 1352-1665 GeV is predicted from $B_q - \bar{B}_q$ mixing by Sahoo et al. [42] and the Drell-Yan process at LHC constrained the mass of $Z'$ boson, mixing angle of $Z - Z'$ and the effective coupling of extra $U(1)$ gauge group [43-45]. The lower bound of $Z'$ mass is set 2.42 TeV [43] for sequential standard model (SSM) and 4.1 TeV [46] for $E_6$ motivated $Z'_\chi$ by the ATLAS collaboration. On the other hand, the lower bound of $Z'$ mass is set 4.5 TeV for sequential standard model (SSM) and 3.9 TeV for superstring-inspired model by the CMS collaboration [44]. The recent Drell-Yan data of LHC reported as $M_{Z'} > 4.4$ TeV by Bandopadhyay et al. when additional $U(1)$ coupling is same as the $SU(2)_L$ coupling [45]. From the study of the mass difference of $B_s$ meson the upper bound of $Z'$ mass is set as 9 TeV in extended standard model [47]. In this work we have considered $Z'$ mass in TeV range.

To study the bounds in the NP couplings generated in our model we have used several observables of other LFV B meson and leptonic decays. Mainly the bound on quark couplings are obtained from $B_s - \bar{B}_s$ mixing and the leptonic couplings are constrained from several experimental upper limits. According to the quark sector level $\Lambda_b$ baryonic and B mesonic decays are induced by same mechanism, so we can independently investigate quark-hadron dynamics with the help of rare decays of baryons, apart from validating the data from the mesonic part.

The paper is organized as follows: In section II we have defined the effective Hamiltonian of the LFV decays in non-universal $Z'$ model. In section III we have discussed the kinematics of the decay, described the hadronic and leptonic helicity amplitudes and structured the observables. We have included $Z'$ contribution in the decays in section IV and performed the numerical analysis in section V. And in section VI we have concluded the findings of our investigation.

## II. Effective Hamiltonian of $b \rightarrow s l_i^+ l_j^-$:

We start to build the effective Hamiltonian with the lepton flavour violating $b \rightarrow s l_i^+ l_j^-$ transition. In the SM, $l_i^+$ and $l_j^-$ leptons are considered of the same flavour $l$ but in our case NP particle $Z'$ will couple with leptons of different family. The Hamiltonian can be written as [19, 25, 28],

$$\mathcal{H}^{eff} = -\frac{G_F \alpha}{2\sqrt{2}\pi} V_{tb} V_{ts}^* \sum_{r=9,10} C'_r O'_r + h.c., \qquad (1)$$

where $G_F$ is the Fermi coupling constant, $\alpha$ electromagnetic coupling constant. The primed parts represent the NP contributions in terms of Wilson Coefficients. Actually the CKM matrix elements $V_{tb}V_{ts}^*$ are included due the virtual effects induced by $t\bar{t}$ contributions. It is to be noted that these LFV decays occur at tree level in our model; therefore the NP is included in such a way that the contributions for $t\bar{t}$ loops are cancelled. Moreover in the SM, there is an electromagnetic operator $O_7$ that contributes in $b \rightarrow sll$ transition but not in LFV part. Non-universal $Z'$ model is basically sensitive for the semileptonic operators including NP contributions in $C'_9$ and $C'_{10}$ [19, 24, 48]. Here,

$$O'_9 = [\bar{s}\gamma_\mu(1-\gamma_5)b][\bar{l}_j\gamma^\mu l_i]$$

$$O'_{10} = [\bar{s}\gamma_\mu(1-\gamma_5)b][\bar{l}_j\gamma^\mu\gamma_5 l_i], \qquad (2)$$

### III. The $\Lambda_b \to \Lambda l_i^+ l_j^-$ decays:

To the best of our knowledge theoretical study of any inclusive decay is easy but their experimental recognition is relatively difficult whereas for the exclusive decays the situation is pretty much opposite. This exclusive decay $\Lambda_b \to \Lambda l_i^+ l_j^-$ is studied in previous section at inclusive level by $b \to s l_i^+ l_j^-$ transition. In this section we have discussed the kinematics of the decay assuming that $\Lambda_b$ is at rest condition whereas $\Lambda$ and dilepton pair travel along positive z -direction and negative z –direction respectively. The momenta of the $\Lambda_b$, $\Lambda$, $l_i$ and $l_j$ are designated by $p$, $k$, $q_i$ and $q_j$ respectively and $s_p$, $s_k$ are the projections of the baryonic spins on z axis in their respective rest frames.

We have considered two kinematic variables as: the four momentum of the dilepton pair $q^\mu = q_i^\mu + q_j^\mu$ and the angle made by $l_i$ lepton with z axis in the dilepton rest frame $\theta_l$. The four momentum of $l_i^+$ is $q_i^\mu$, where $l_j^-$ has four momentum $q_j^\mu$. The four momentums can be defined as below,

$$q_i^\mu|_{2l} = (E_2, -|q_{2l}|\sin\theta_l, 0, -|q_{2l}|\cos\theta_l), \quad q_j^\mu|_{2l} = (E_1, |q_{2l}|\sin\theta_l, 0, |q_{2l}|\cos\theta_l). \qquad (3)$$

Where,

$$|q_{2l}| = \frac{\sqrt{\lambda(q^2, m_i^2, m_j^2)}}{2\sqrt{q^2}}, \quad E_1 = \frac{q^2 + m_i^2 - m_j^2}{2\sqrt{q^2}}, \quad E_2 = \frac{q^2 - m_i^2 + m_j^2}{2\sqrt{q^2}}, \qquad (4)$$

Here, the physical range of momentum transferred term $q^2$ is defined as,
$$(m_j^2 + m_i^2) \leq q^2 \leq (m_{\Lambda_b} - m_\Lambda)^2. \qquad (5)$$

The decay amplitudes can be written in terms of hadronic and leptonic helicity amplitudes which is given as below,

$$\mathcal{M}^{\lambda_j,\lambda_i}(s_p, s_k) = -\frac{G_F \alpha}{2\sqrt{2}\pi} V_{tb} V_{ts}^* \sum_{h=L,R}\left[\sum_\lambda \eta_\lambda H_{VA,\lambda}^{h,s_p,s_k} L_{h,\lambda}^{\lambda_2,\lambda_1}\right], \qquad (6)$$

where $H_{VA,\lambda}^{h,s_p,s_k}$ are vector and axial-vector related hadronic helicity amplitudes and $L_{h,\lambda}^{\lambda_j,\lambda_i}$ are the leptonic helicity amplitudes. In the above eq. (6), $h = L, R$ represent the chiralities of the lepton current, $\lambda = t, \pm 1, 0$ represent the helicity states of the virtual gauge boson that decays into dilepton pair, $\lambda_{i,j}$ represent the helicities of the leptons and $\eta_t = 1$, $\eta_{\pm 1,0} = -1$. The hadronic helicity amplitudes can be represented in terms of the Wilson Coefficients and form factors which are defined in the ref. [49]. In our work we are interested in the transversity amplitudes [24, 50] $A^h_{\perp(\parallel)_1}, A^h_{\perp(\parallel)_0}$ and $A_{\perp(\parallel)t}$. These expressions are defined in Appendix A. The $\Lambda_b \to \Lambda$ transition matrix elements are parametrized correlating the operators of eq. (2) in terms of $q^2$ dependent form factors [51] $f^V_{0,t,\perp}$ and $f^A_{0,t,\perp}$ and their numerical values are taken from the ref. [52]. For completeness of the paper the detail about the form factors are collected in Appendix B.

The leptonic helicity amplitudes can be defined as below and the details are discussed in Appendix C,

$$L_{L(R),\lambda}^{\lambda_j,\lambda_i} = \bar{\epsilon}^\mu(\lambda)\langle \bar{l}_j(\lambda_j)l_i(\lambda_i)|\bar{l}_j\gamma_\mu(1 \mp \gamma_5)l_i|0\rangle, \tag{7}$$

where $\epsilon^\mu$ represents the polarization vector of the virtual gauge boson which is involved in the decays of dilepton pair [49]. Proceeding with all these calculations we have found the expression of two fold differential branching ratio for this decay as,

$$\frac{d^2\mathcal{B}}{dq^2 d\cos\theta_l} = \frac{3}{2}(K_{1ss}\sin^2\theta_l + K_{1cc}\cos^2\theta_l + K_{1c}\cos\theta_l) \tag{8}$$

The angular coefficients $K_{1ss,1cc,1c}$ can be written in terms of the transversity amplitudes which are given as below,

$$K_{1ss} = \frac{1}{4}\left(2|A_{\parallel 0}^R|^2 + |A_{\parallel 1}^R|^2 + 2|A_{\perp 0}^R|^2 + |A_{\perp 1}^R|^2 + \{R \leftrightarrow L\}\right)$$
$$- \frac{(m_i^2 + m_j^2)}{2q^2}\left[\left(|A_{\parallel 0}^R|^2 + |A_{\perp 0}^R|^2 + \{R \leftrightarrow L\}\right) - (|A_{\perp t}|^2 + \{\perp \leftrightarrow \parallel\})\right]$$
$$+ \frac{(m_i^2 m_j^2)}{q^2}[2Re(A_{\perp 0}^R A_{\perp 0}^{*L} + A_{\perp 1}^R A_{\perp 1}^{*L} + \{\perp \leftrightarrow \parallel\})]$$
$$- \frac{(m_i^2 - m_j^2)^2}{4q^4}\left[\left(|A_{\parallel 1}^R|^2 + |A_{\perp 1}^R|^2 + \{R \leftrightarrow L\}\right) + 2\left(|A_{\parallel t}|^2 + |A_{\perp t}|^2\right)\right], \tag{9}$$

$$K_{1cc} = \frac{1}{2}\left(|A_{\parallel 1}^R|^2 + |A_{\perp 1}^R|^2 + \{R \leftrightarrow L\}\right)$$
$$+ \frac{(m_i^2 + m_j^2)}{2q^2}\left[\left(|A_{\parallel 0}^R|^2 + |A_{\perp 0}^R|^2 - |A_{\parallel 1}^R|^2 - |A_{\perp 1}^R|^2 + \{R \leftrightarrow L\}\right)\right.$$
$$\left. + \left(|A_{\parallel t}|^2 + |A_{\perp t}|^2\right)\right] + \frac{(m_i^2 m_j^2)}{q^2}[2Re(A_{\perp 0}^R A_{\perp 0}^{*L} + A_{\perp 1}^R A_{\perp 1}^{*L} + \{\perp \leftrightarrow \parallel\})]$$
$$- \frac{(m_i^2 - m_j^2)^2}{4q^4}\left[\left(|A_{\parallel 0}^R|^2 + |A_{\perp 0}^R|^2 + \{R \leftrightarrow L\}\right) + \left(|A_{\parallel t}|^2 + |A_{\perp t}|^2\right)\right], \tag{10}$$

$$K_{1c} = -\beta\beta'\left(A_{\perp 1}^R A_{\parallel 1}^{*R} - \{R \leftrightarrow L\}\right) + \beta\beta'\frac{(m_i^2 - m_j^2)}{q^2}Re(A_{\parallel 0}^L A_{\parallel t}^* + A_{\perp 0}^L A_{\perp t}^*), \tag{11}$$

where, $\beta = \sqrt{1 - \frac{(m_i+m_j)^2}{q^2}}$ and $\beta' = \sqrt{1 - \frac{(m_i-m_j)^2}{q^2}}$. \qquad (12)

We have informed previously that the paper is mainly concentrated on the differential branching fractions and forward backward asymmetries of $\Lambda_b \to \Lambda l_i^+ l_j^-$ decays calculated from the double differential distribution of eq. (8). Integrating eq. (8) with respect to $\cos\theta_l$, we have obtained as below:

$$\frac{d\mathcal{B}}{dq^2} = 2K_{1ss} + K_{1cc}. \tag{13}$$

Another powerful tool for searching new physics in LFV decays is the forward-backward asymmetry which is given by,

$$A_{FB}(q^2) = \frac{\int_0^1 \frac{d^2\mathcal{B}}{dq^2 d\cos\theta_l} d\cos\theta_l - \int_{-1}^0 \frac{d^2\mathcal{B}}{dq^2 d\cos\theta_l} d\cos\theta_l}{\int_0^1 \frac{d^2\mathcal{B}}{dq^2 d\cos\theta_l} d\cos\theta_l + \int_{-1}^0 \frac{d^2\mathcal{B}}{dq^2 d\cos\theta_l} d\cos\theta_l}. \tag{14}$$

Using all the above equations we have obtained the following form as:

$$A_{FB}(q^2) = \frac{3}{2}\frac{K_{1c}}{(K_{1ss} + K_{1cc})}. \tag{15}$$

### IV. Contribution of $Z'$ boson in $\Lambda_b \to \Lambda l_i^+ l_j^-$ decays:

Non-universal $Z'$ model is one of the most simplified models in which NP effects originate from a heavy new gauge boson $Z'$ with various couplings to quarks and charged lepton pairs. In this $Z'$ model, one extra $U(1)'$ gauge group is associated with the SM gauge group [36, 53]. The characteristic of the $Z'$ couplings with fermions explains the FCNC transitions to the tree level successfully. Some theories have assumed the flavour universal $Z'$ couplings considering the diagonal element even in the presence of flavour mixing of fermions by the GIM mechanism [36]. However, for construction of different families we have to consider the family non-universal $Z'$ couplings in BSMs. Chaudhuri, Hockney and Lykken [54-56] have predicted that the third generation of quark couples differently to $Z'$ from the other two families as well as all three lepton generations also couples differently to the $Z'$ boson. Here we have assumed that our $Z'$ couplings are diagonal and non-universal in nature.

With the extension towards BSM the $U(1)'$ currents represented as,

$$J_\mu = \sum_{i,j} \bar{\psi}_j \gamma_\mu \left[\epsilon_{\psi_{L_{ij}}} P_L + \epsilon_{\psi_{R_{ij}}} P_R\right] \psi_i. \tag{16}$$

Here the sum extends over all fermions $\psi_{i,j}$ and $\epsilon_{\psi_{R,L_{ij}}}$ denote the chiral couplings of the newly originated gauge boson. We have also assumed that our $Z'$ couplings are diagonal and non-universal in nature. Generally FCNCs form in both LH and RH sectors at the tree level. So we can write as, $B_{ij}^{\psi_L} \equiv \left(V_L^\psi \epsilon_{\psi_L} V_L^{\psi\dagger}\right)_{ij}$, $B_{ij}^{\psi_R} \equiv \left(V_R^\psi \epsilon_{\psi_R} V_R^{\psi\dagger}\right)_{ij}$. The $Z'\bar{b}q$ couplings can be generated as

$$\mathcal{L}_{FCNC}^{Z'} = -g'\left(B_{sb}^L \bar{s}_L \gamma_\mu b_L + B_{sb}^R \bar{s}_R \gamma_\mu b_R\right)Z'^\mu + h.c., \tag{17}$$

where $g'$ is the gauge coupling associated with the $U(1)'$ group and the effective Hamiltonian becomes as,

$$H_{eff}^{Z'} = \frac{8G_F}{\sqrt{2}} \left( \rho_{sb}^L \bar{s}_L \gamma_\mu b_L + \rho_{sb}^R \bar{s}_R \gamma_\mu b_R \right) \left( \rho_{l_i l_j}^L \bar{l}_{j_L} \gamma_\mu l_{i_L} + \rho_{l_i l_j}^R \bar{l}_{j_R} \gamma_\mu l_{i_R} \right), \tag{18}$$

where $\rho_{l_i l_j}^{L,R} \equiv \frac{g' M_Z}{g M_{Z'}} B_{l_i l_j}^{L,R}$, $g'$ and $g$ are the gauge couplings of $Z'$ and $Z$ bosons (where $g = \frac{e}{\sin\theta_W \cos\theta_W}$) respectively. Here we need to consider some simplifications: (i) we have neglected kinetic mixing as it represents the redefinition of unknown couplings, (ii) we have also neglected $Z - Z'$ mixing [36, 57-60] (as the mixing angle is constrained as less than $10^{-3}$ by Bandyopadhyay et al. [45] and recently at LHC it was found as of the order $10^{-4}$ by Bobovnikov et al. [61]) for its very small mixing angle, (iii) there are no crucial effects of renormalization group (RG) evolution between the mass of W boson ($M_W$) and the mass of $Z'(M_{Z'})$ scales, (iv) we consider the remarkable contribution of the flavour-off-diagonal left-handed couplings of quarks in the flavour changing $b - s - Z'$ part [62-66]. Along with these simplifications the LHC Drell-Yan data have constrained the following parameters recently: the mass of $Z'$ boson, the $Z - Z'$ mixing angle and the extra $U(1)'$ gauge coupling. Not only the $Z - Z'$ mixing angle but also the $Z'$ mass have constrained as $M_{Z'} < 4.4$ TeV by Bandyopadhyay et al. [45]. The value of $\left|\frac{g'}{g}\right|$ is not determined yet. But we can expect that $\left|\frac{g'}{g}\right| \sim 1$ as both $U(1)$ gauge groups generate from same origin of some grand unified theory and $\left|\frac{M_Z}{M_{Z'}}\right| \sim 0.1$ for TeV-scale $Z'$. All four experiments of LEP have also suggested the existence of $Z'$ boson with the same fermionic coupings as that of the SM $Z$ boson. If $|\rho_{sb}^L| \sim |V_{tb} V_{ts}^*|$, then the order of $B_{sb}^L$ will be of $10^{-3}$. Considering all the simplifications for the non-universal couplings of $Z'$, the effective Hamiltonian including the NP part for $b \to s l_i^+ l_j^-$ becomes

$$H_{eff}^{Z'} = -\frac{2G_F}{\sqrt{2}\pi} V_{tb} V_{ts}^* \left[ \frac{B_{sb}^L B_{l_i l_j}^L}{V_{tb} V_{ts}^*} (\bar{s}b)_{V-A} (\bar{l}_j l_i)_{V-A} - \frac{B_{sb}^L B_{l_i l_j}^R}{V_{tb} V_{ts}^*} (\bar{s}b)_{V-A} (\bar{l}_j l_i)_{V+A} \right] + h.c., \tag{19}$$

where $B_{sb}^L$ is the left-handed coupling of $Z'$ boson with quarks, $B_{l_i l_j}^L$ and $B_{l_i l_j}^R$ are the left-handed and right-handed couplings with the leptons respectively. The coupling parameter consists of a NP weak phase term which is related as, $B_{sb}^L = |B_{sb}^L| e^{-i\varphi_s^l}$.

With the assistance of eq. (1) and eq. (2), we have included the NP terms as follows:

$$C_9' = \frac{4\pi B_{sb}^L}{\alpha V_{tb} V_{ts}^*} \left( B_{l_i l_j}^L + B_{l_i l_j}^R \right),$$
$$C_{10}' = \frac{4\pi B_{sb}^L}{\alpha V_{tb} V_{ts}^*} \left( B_{l_i l_j}^L - B_{l_i l_j}^R \right). \tag{20}$$

With all above considerations we have studied the observables defined in previous section in the light of new physics.

## V. Numerical Analysis:

In this work we have studied differential branching ratio and forward backward asymmetry for the lepton flavour violating decay mode $\Lambda_b \to \Lambda l_i^+ l_j^-$ in the framework of non-universal $Z'$ model. In this work we are mainly focussing on the terms $C_9'$ and $C_{10}'$ in which NP affects dominantly and here it is essential to fix the $b-s-Z'$ coupling as well as the NP weak phase $\varphi_s^l$ which are constrained from $B_s - \bar{B}_s$ mixing data of UTfit Collaboration [67-73] and various inclusive and exclusive decays. The numerical values of these NP couplings are recorded below in Table-1. The first two scenarios are taken from refs. [33, 48]. The third scenario is constrained from the recent values of the mass differences updated in ref [47]. The constrained values are taken from our previous work [74]. Other input parameters are taken from Appendix D [75]. Using all these values in the expressions we have shown the variation of the observables with leptonic couplings within the kinematically accessible physical range of $q^2$.

**Table-1: Numerical values of coupling parameters**

| Scenarios | $\lvert B_{sb} \rvert \times 10^{-3}$ | $\varphi_s^l$ (in degree) |
|---|---|---|
| $S_1$ | $(1.09 \pm 0.22)$ | $(-72 \pm 7)°$ |
| $S_2$ | $(2.20 \pm 0.15)$ | $(-82 \pm 4)°$ |
| $S_3$ | $0 \leq \lvert B_{sb}^L \rvert \leq 1.539 \times 10^{-3}$ | For $0° \leq \varphi_s^L \leq 180°$ |

To magnify the influence of NP in the observables we have considered the maximum values of the NP couplings. According to the ranges of three scenarios given in Table-1 we have fixed three sets of values of coupling parameters which are given below.

**For scenario 1:** Taken from the range of $S_1$ in Table-1 the maximum magnitude of the NP coupling parameter $\lvert B_{sb} \rvert$ and NP weak phase angle $\varphi_s^l$ are set as $\lvert B_{sb} \rvert = (1.31 \times 10^{-3})$ and $\varphi_s^l = -65°$.

**For scenario 2:** Again in accordance with the limit for $S_2$ in Table-1 the enhanced effect of NP couplings are set as $\lvert B_{sb} \rvert = (2.35 \times 10^{-3})$ and $\varphi_s^l = -78°$.

**For scenario 3:** The maximum contributions of NP couplings for 3$^{\text{rd}}$ scenario are set as $\lvert B_{sb} \rvert = (1.539 \times 10^{-3})$ and $\varphi_s^l = 180°$.

With all these numerical data we have made a start of our investigation. According to eq. (20) the left (right) handed leptonic couplings $B_{l_i l_j}^{L(R)}$ represent lepton flavour violating nature of the decays. Now we need to find the bounds on these couplings utilising the experimental upper limits constrained by LHCb [3, 75-78] on the LFV decays.

In order to study the limits of LFV leptonic couplings of $\Lambda_b \to \Lambda \mu^+ e^-$ decay having inclusive transition as $b \to s \mu^+ e^-$, we have used the following experimental bounds of branching ratios of same quark transition set by LHCb [75-78]

$$\mathcal{B}(B_s \to \mu^\pm e^\mp) < (5.4 \times 10^{-9}), \text{ with 95\% C. L.}$$
$$\mathcal{B}(\mu \to 3e) < 10^{-12},$$
$$\Delta\mathcal{B}(\mu \to e\upsilon_\mu \bar{\upsilon}_e) \leq (4 \times 10^{-5}),$$
$$\mathcal{B}(B^+ \to K^+\mu^+ e^-) < (8.8 \times 10^{-9}), \text{ with 95\% C. L.} \quad (21)$$

We have formulated the expressions for these observables and plotted them within the mentioned experimental bound. The allowed region provides the possible bound of the NP couplings. The expression for the branching ratios for LFV decays are followed from the ref. [19] and for $B_s \to l_i^+ l_j^-$ decays it is taken as,

$$\mathcal{B}(B_s \to l_i^+ l_j^-)$$
$$= \frac{\tau_B (m_i + m_j)^2 \alpha^2 G_F^2}{2^5 \pi^3 m_B} |V_{tb} V_{ts}^*|^2 \sqrt{m_i^4 + m_j^4 + m_B^4 - 2(m_i^2 m_j^2 + m_i^2 m_B^2 + m_j^2 m_B^2)} (|C_9|^2 + |C_{10}|^2), \quad (22)$$

The branching ratio expressions for other LFV decays are also structured as following:

$$\mathcal{B}(B^+ \to K^+ l_i^+ l_j^-) = 10^{-9} \left( a_{l_i^+ l_j^-} |C_9|^2 + b_{l_i^+ l_j^-} |C_{10}|^2 \right), \quad (23)$$

The values of $a_{l_i^+ l_j^-}$ and $b_{l_i^+ l_j^-}$ are taken from the ref. [19].

$$\mathcal{B}(l_i \to 3l_j) = \frac{\tau_{l_i} m_{l_i}^5}{1536\pi^3 m_{Z'}^4} \left[ 2\left( \left|B_{l_i l_j}^L B_{l_j l_j}^L\right|^2 + \left|B_{l_i l_j}^R B_{l_j l_j}^R\right|^2 \right) + \left|B_{l_i l_j}^L B_{l_j l_j}^R\right|^2 + \left|B_{l_i l_j}^R B_{l_j l_j}^L\right|^2 \right], \quad (24)$$

Here $\Delta\mathcal{B}\left(l_i \to l_j \upsilon_{l_i} \bar{\upsilon}_{l_j}\right)$ is represented as,

$$\Delta\mathcal{B}\left(l_i \to l_j \upsilon_{l_i} \bar{\upsilon}_{l_j}\right) = \left[\mathcal{B}\left(l_i \to l_j \upsilon_{l_i} \bar{\upsilon}_{l_j}\right)\right]_{exp} - \left[\mathcal{B}\left(l_i \to l_j \upsilon_{l_i} \bar{\upsilon}_{l_j}\right)\right]_{SM} \quad (25)$$

We have plotted the observables in Fig. 1 with respect to their NP couplings within the experimental upper limits mentioned at eq. (21). In Fig. 1 the red contour represents the parameter space for $\mathcal{B}(B^+ \to K^+\mu^+ e^-)$, green contour represents for $\mathcal{B}(B_s \to \mu^+ e^-)$ and yellow region is for the LFV decay $\mathcal{B}(\mu \to 3e)$. The common portion is considered as the allowed region for the NP couplings $B_{\mu e}^L$ and $B_{\mu e}^R$. The upper limit of $\Delta\mathcal{B}\left(l_i \to l_j \upsilon_{l_i} \bar{\upsilon}_{l_j}\right)$ bounded $B_{\mu e}^L$ within the range that obtained from Fig. 1.

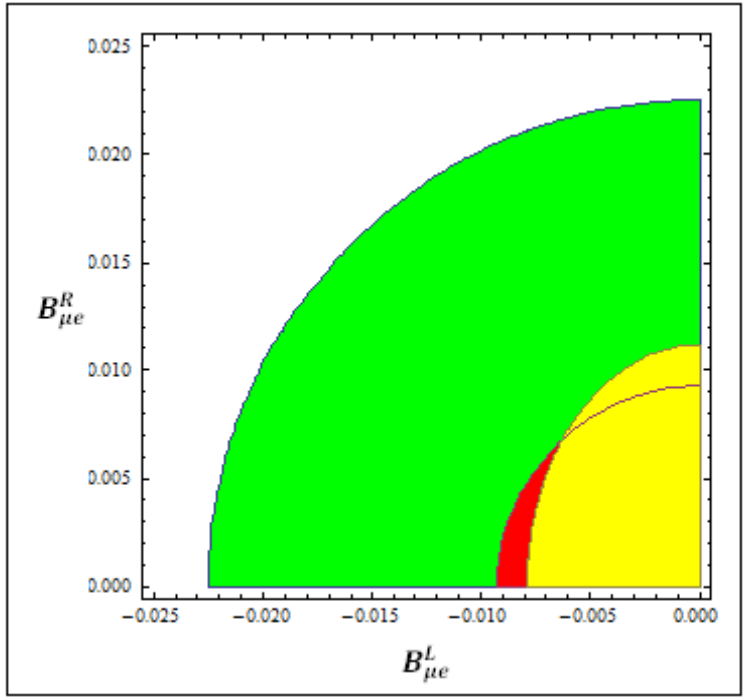

**Fig. 1: The parameter space allowed by the experimental upper limits mentioned at eq. (21)**

According to the Fig. 1 we have set the bounds for the leptonic couplings. Here we can see that $B_{\mu e}^L = -0.0079$ when $B_{\mu e}^R = 0$ whereas $B_{\mu e}^L$ set to zero when $B_{\mu e}^R = 0.0092$. In terms of these couplings we have formulated $S_{l_i l_j}$ and $D_{l_i l_j}$ as below:

$$S_{l_i l_j} = \left(B_{l_i l_j}^L + B_{l_i l_j}^R\right), \text{ and } D_{l_i l_j} = \left(B_{l_i l_j}^L - B_{l_i l_j}^R\right). \tag{26}$$

Another consideration is taken in our model as: $C_9' = -C_{10}'$ (this consideration provided many fruitful results in these references [24, 50, 79-82] also). Therefore, we have studied within the parameter space from $(-0.0079)$ to $0.0079$. To enhance the contribution of our model in this decay we have taken the maximum contribution of the leptonic couplings as $0.0079$. Using all the values of NP couplings we have varied differential branching ratio within allowed kinematic region of $q^2$ in Fig. 2a, Fig. 2b and Fig. 2c for scenario 1, scenario 2 and scenario 3 respectively. The blue line shows the variation of differential branching ratio taking the central values of the form factors and input parameters whereas the red and the black line show the maximum and minimum variation of the observable in accordance with the uncertainties of all the form factors and input parameters. From this figure we can observe that the value of differential branching ratio increases with increment of $q^2$. Another thing is that the differential branching ratio has maximum value for scenario 2 which indicates the sensitivity of NP on $\Lambda_b \to \Lambda \mu^+ e^-$ decay.

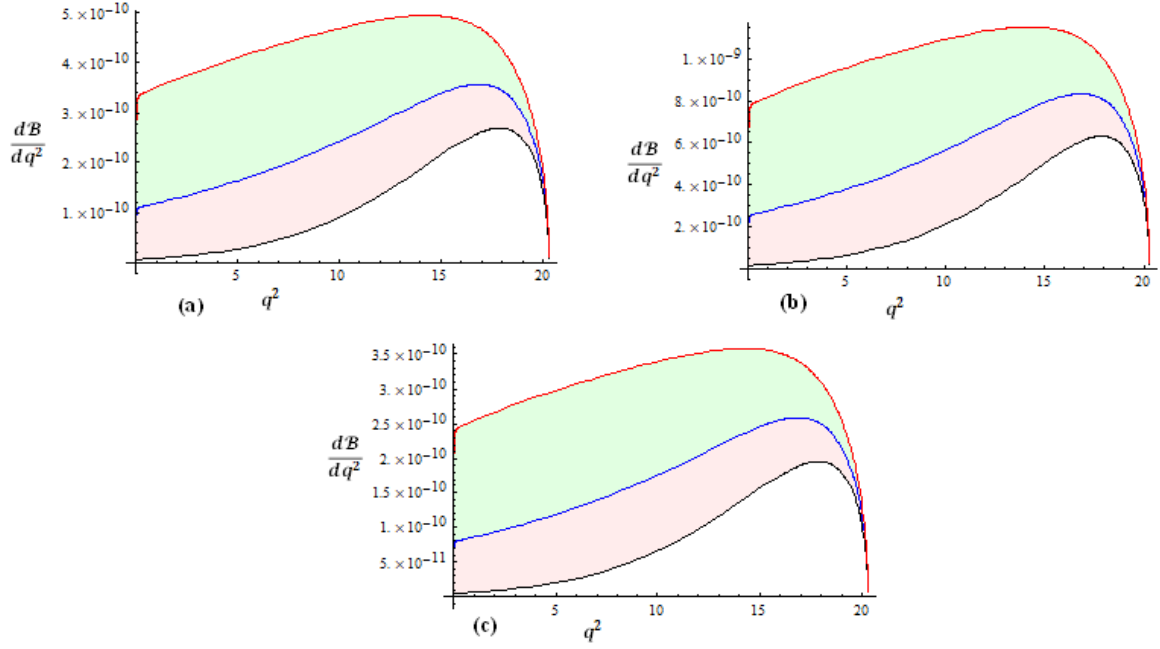

**Fig. 2: Variation of differential branching ratio for $\Lambda_b \to \Lambda \mu^+ e^-$ within allowed kinematic region of $q^2$ using the bound of NP couplings for (a) scenario 1, (b) scenario 2 and (c) scenario 3**

Similarly, we have plotted the variation of forward backward asymmetry within the same parameter space in Fig. 3a, Fig. 3b and Fig. 3c for scenario 1, scenario 2 and scenario 3 respectively. The interpretation of colour bands of the figures is discussed previously. According to the figures, it is observed that at low $q^2$ region the value of the observable is slightly negative then it increases gradually. Here $A_{FB}$ zero crossing is present at $q^2 = 0.6$ GeV$^2$ for 1st and 3rd scenario and $q^2 = 0.4$ GeV$^2$ for 2nd scenario which are shown with magnified view at Fig. 3d. Here, we can observe that the zero crossing values for 1st and 3rd scenarios are almost same and larger the value of NP coupling $|B_{sb}|$ smaller the value of zero crossing. From $q^2 = 10$ GeV$^2$ $A_{FB}$ increases significantly and then suddenly drops at high momentum region. Here, we should note that forward backward asymmetry increases for scenario 2 due to more contribution of NP. The calculated values of differential branching ratio and forward backward asymmetry with the upper bounds of the couplings are recorded in Table-2.

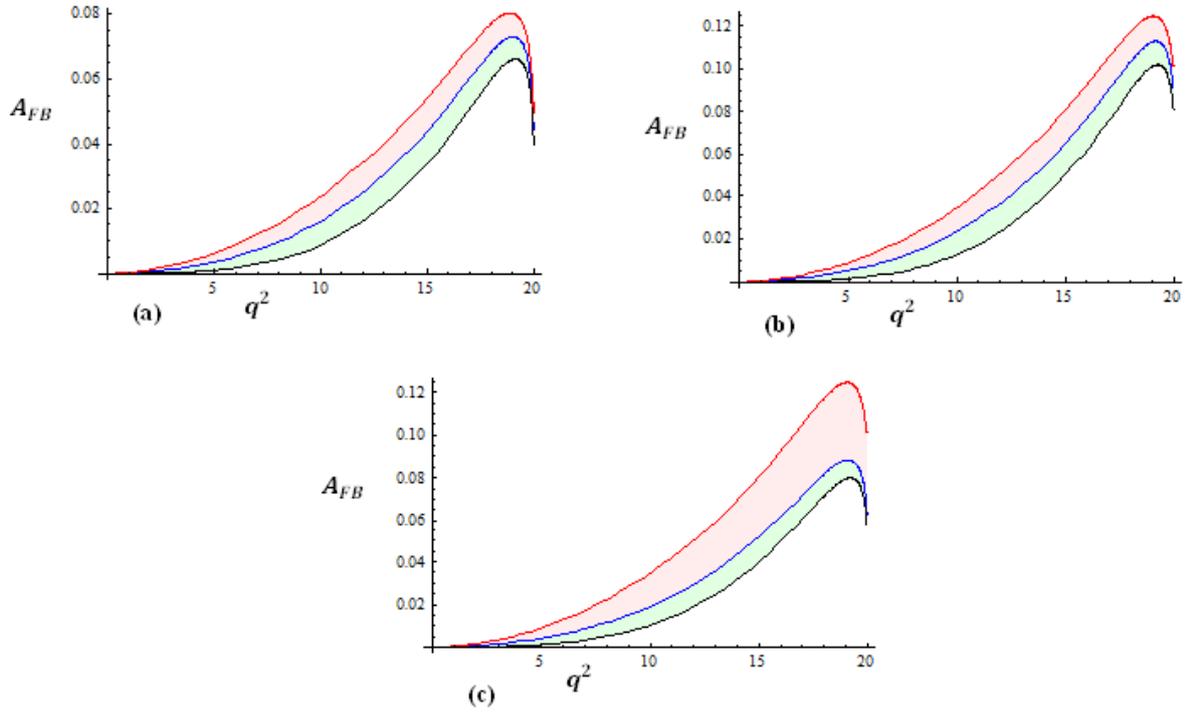

**Fig. 3: Variation of lepton side forward backward asymmetry for $\Lambda_b \to \Lambda \mu^+ e^-$ within allowed kinematic region of $q^2$ using the bound of NP couplings for (a) scenario 1, (b) scenario 2 and (c) scenario 3**

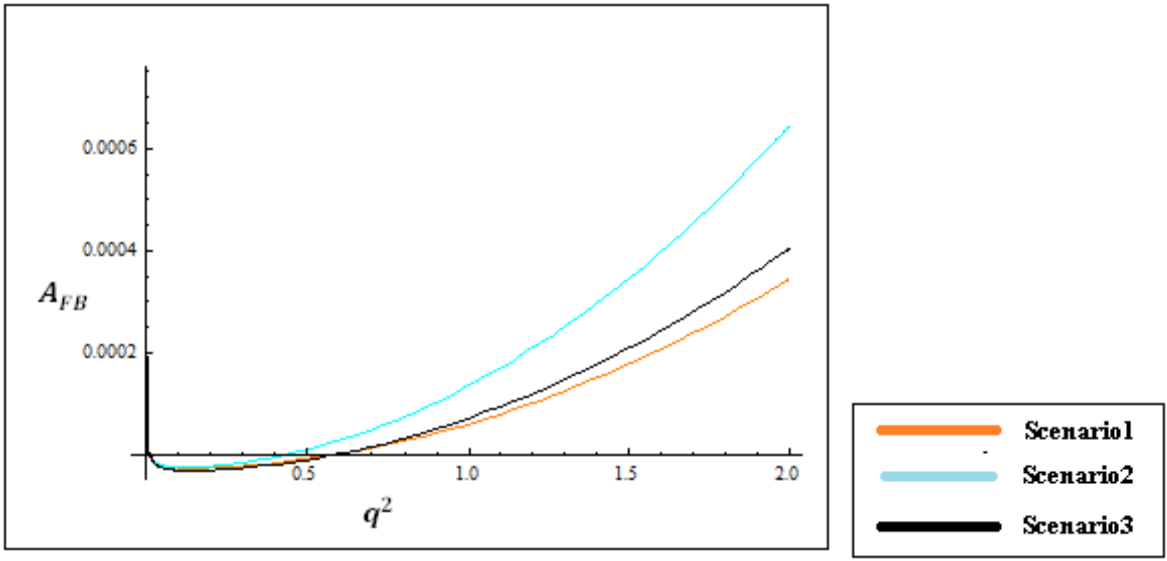

**Fig. 3d: Variation of lepton side forward backward asymmetry for $\Lambda_b \to \Lambda \mu^+ e^-$ in low $q^2$ region to locate the position of zero crossing**

**Table-2: Predicted values of differential branching ratios and forward backward asymmetries for $\Lambda_b \to \Lambda \mu^+ e^-$ decay in 1st, 2nd and 3rd scenarios**

| Kinematic region ($q^2$) (in GeV²) | For $\Lambda_b \to \Lambda \mu^+ e^-$ | | |
|---|---|---|---|
| | $\dfrac{d\mathcal{B}}{dq^2}$ | | |
| | 1st Scenario | 2nd Scenario | 3rd Scenario |
| In $q^2 = 6$ | $(1.67 \pm 1.20) \times 10^{-10}$ | $(5.36 \pm 3.52) \times 10^{-10}$ | $(2.30 \pm 1.54) \times 10^{-10}$ |
| In $q^2 = 12$ | $(2.24 \pm 1.22) \times 10^{-10}$ | $(7.20 \pm 2.10) \times 10^{-10}$ | $(3.09 \pm 1.46) \times 10^{-10}$ |
| In $q^2 = 18$ | $(2.53 \pm 0.57) \times 10^{-10}$ | $(8.14 \pm 1.83) \times 10^{-10}$ | $(3.49 \pm 0.78) \times 10^{-10}$ |
| | $A_{FB}$ | | |
| | 1st Scenario | 2nd Scenario | 3rd Scenario |
| In $q^2 = 6$ | $(0.005 \pm 0.003)$ | $(0.007 \pm 0.004)$ | $(0.006 \pm 0.004)$ |
| In $q^2 = 12$ | $(0.024 \pm 0.009)$ | $(0.035 \pm 0.013)$ | $(0.028 \pm 0.012)$ |
| In $q^2 = 18$ | $(0.066 \pm 0.009)$ | $(0.101 \pm 0.014)$ | $(0.079 \pm 0.012)$ |

To study the $\Lambda_b \to \Lambda \tau^+ \mu^-$ decay we need to set the upper limit of leptonic couplings $B_{\tau\mu}^L$ and $B_{\tau\mu}^R$. We have used the experimental bounds of some LFV decays. Previously BaBaR collaboration has obtained the upper limit of the branching ratio for $B^0 \to \tau^\pm \mu^\mp$ as $(2.2 \times 10^{-5})$ at 90% C.L. but has not obtained any data for $B_s \to \tau^\pm \mu^\mp$. First time this bound was set by LHCb in the ref [3]. Other experimental limits that are used in this work are as follows [3, 75, 83],

$$\mathcal{B}(B_s \to \tau^\pm \mu^\mp) < (4.2 \times 10^{-5}), \text{ with 95\% C. L.}$$
$$\mathcal{B}(\tau \to 3\mu) < (2.1 \times 10^{-8}),$$
$$\mathcal{B}(B^+ \to K^+ \tau^\pm \mu^\mp) < (4.8 \times 10^{-5}), \text{ with 90\% C. L.} \quad (27)$$

As previous we have plotted the above observables mentioned at eq. (27) with respect to the NP couplings in the Fig. 4 where red portion represents the branching ratio for $B^+ \to K^+ \tau^\pm \mu^\mp$ decay, green portion for $\tau \to 3\mu$ decay and blue portion represents $B_s \to \tau^\pm \mu^\mp$ decay. According to the figure the common portion is the allowed region. Along with the previous considerations we have enhanced the contribution of the NP leptonic couplings by fixing it at 0.11.

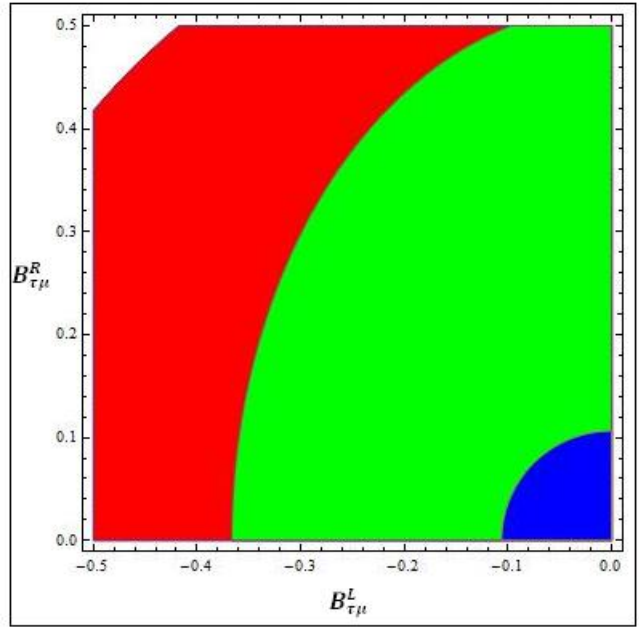

**Fig. 4: The parameter space allowed by the experimental upper limits mentioned at eq. (27)**

At first we have studied the variation of differential branching ratio for $\Lambda_b \rightarrow \Lambda \tau^+ \mu^-$ decay within the whole allowable kinematic region in Fig. 5a, Fig. 5b and Fig. 5c for scenario 1, scenario 2 and scenario 3 respectively. Here we can observe that differential branching ratio for 2$^{nd}$ scenario increases sharply in compare to the 1$^{st}$ and 3$^{rd}$ scenario and the observable attained largest value for the 2$^{nd}$ one. So we can say that the observable is sensitive to our NP model.

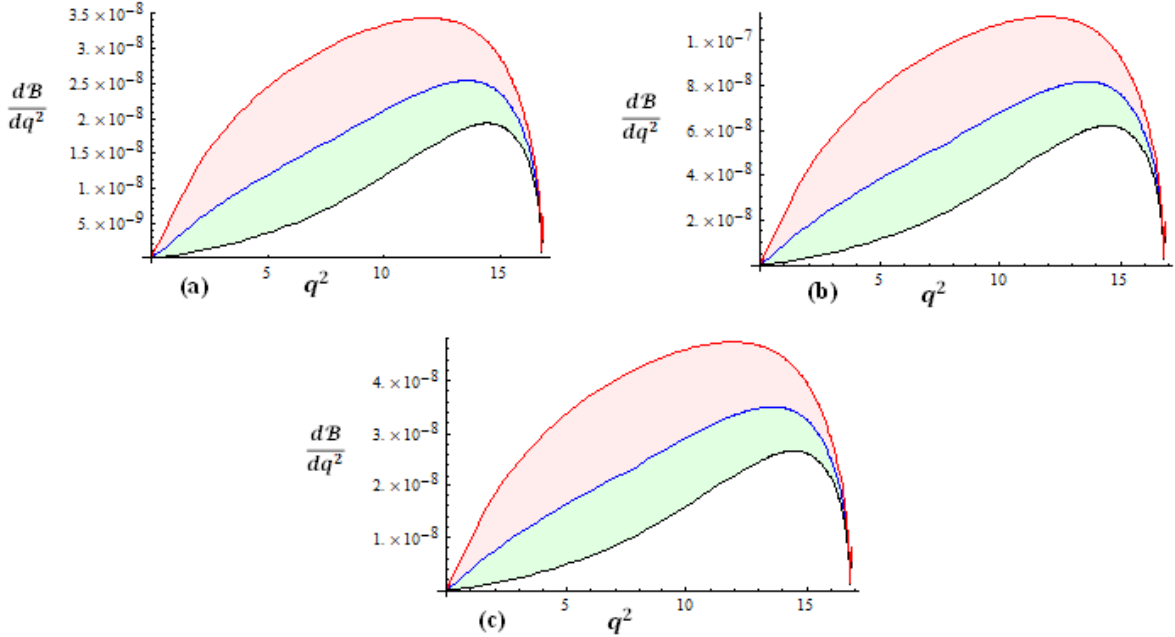

**Fig. 5: Variation of differential branching ratio for $\Lambda_b \rightarrow \Lambda \tau^+ \mu^-$ within allowed kinematic region of $q^2$ using the bound of NP couplings for (a) scenario 1, (b) scenario 2 and (c) scenario 3**

Secondly we have varied the forward backward asymmetry for this LFV decay with respect to the total kinematic range in Fig. 6a, Fig. 6b and Fig. 6c for scenario 1, scenario 2 and scenario 3 respectively. Here we have seen that $A_{FB}$ increases more sharply for 2$^{nd}$ scenario than for the 1$^{st}$ and 3$^{rd}$ scenarios. From about $q^2 = 10$ GeV$^2$ this value enhances significantly. The zero crossing for 2$^{nd}$ scenario is nearer to the origin than the other scenarios. Fig. 6d shows the variation of $A_{FB}$ with respect to $q^2$ for low $q^2$ region to show the zero crossing of observable. The calculated values of differential branching ratio and $A_{FB}$ for $\Lambda_b \to \Lambda \tau^+ \mu^-$ decay are encapsulated in Table 3.

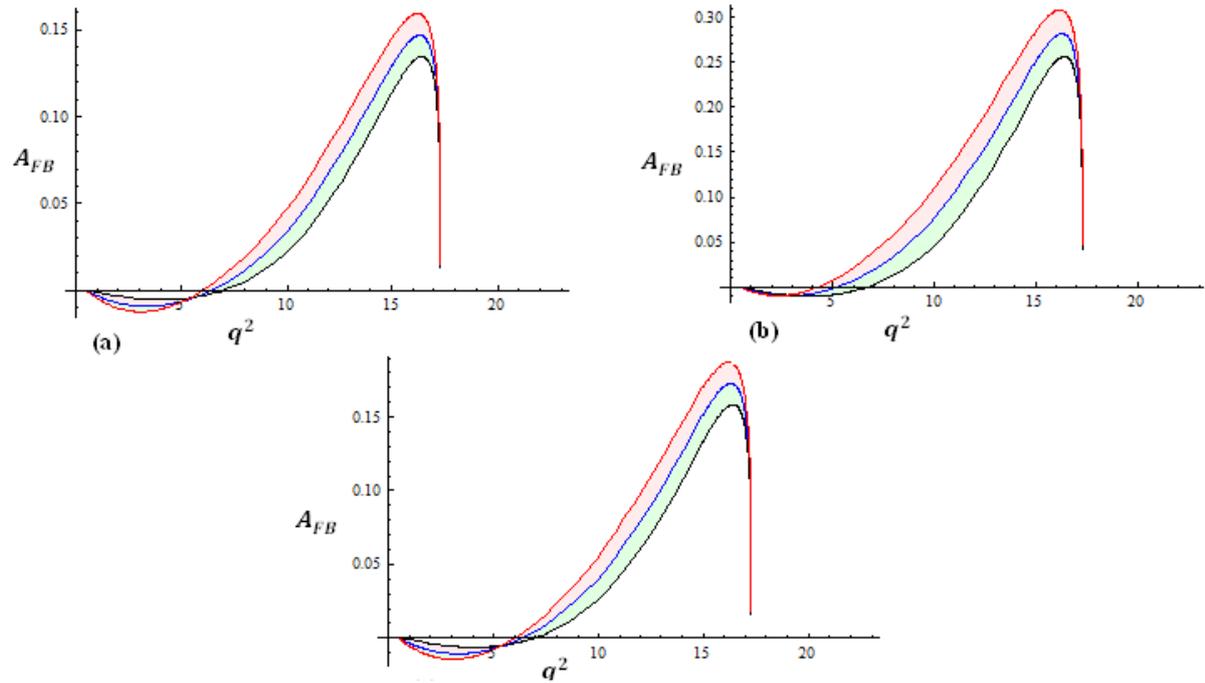

**Fig. 6: Variation of lepton side forward backward asymmetry for $\Lambda_b \to \Lambda \tau^+ \mu^-$ within allowed kinematic region of $q^2$ using the bound of NP couplings for (a) scenario 1, (b) scenario 2 and (c) scenario 3**

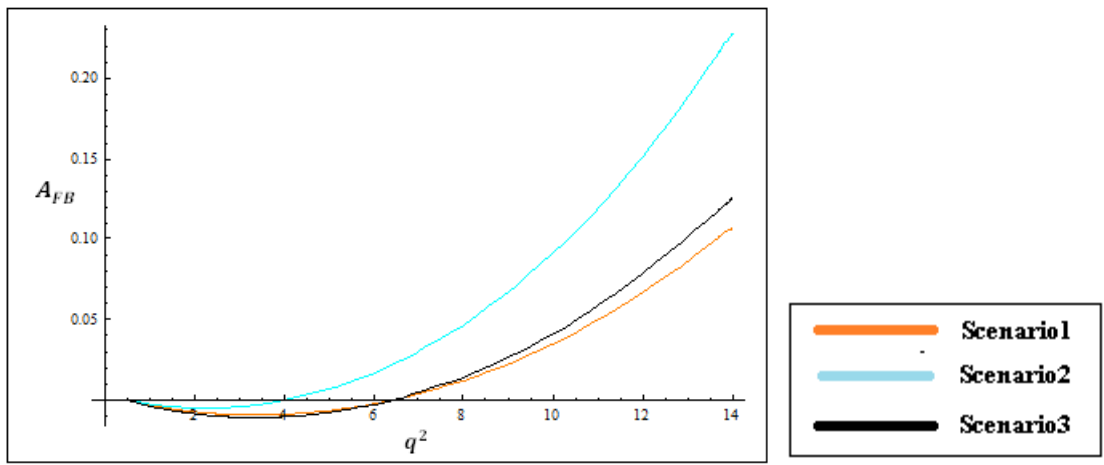

**Fig. 6d: Variation of lepton side forward backward asymmetry for $\Lambda_b \to \Lambda \tau^+ \mu^-$ in low $q^2$ region to locate the position of zero crossing**

**Table-3: Predicted values of differential branching ratios and forward backward asymmetries for $\Lambda_b \to \Lambda \tau^+ \mu^-$ decay in 1st, 2nd and 3rd scenarios**

| Kinematic region ($q^2$) (in GeV$^2$) | For $\Lambda_b \to \Lambda \tau^+ \mu^-$ | | |
|---|---|---|---|
| | $\dfrac{d\mathcal{B}}{dq^2}$ | | |
| | 1st Scenario | 2nd Scenario | 3rd Scenario |
| In $q^2 = 6$ | $(8.45 \pm 5.11) \times 10^{-9}$ | $(2.72 \pm 1.15) \times 10^{-8}$ | $(1.16 \pm 0.98) \times 10^{-8}$ |
| In $q^2 = 12$ | $(2.01 \pm 1.14) \times 10^{-8}$ | $(6.46 \pm 2.68) \times 10^{-8}$ | $(2.77 \pm 1.58) \times 10^{-8}$ |
| In $q^2 = 18$ | $(2.51 \pm 0.57) \times 10^{-8}$ | $(8.07 \pm 1.84) \times 10^{-8}$ | $(3.46 \pm 0.78) \times 10^{-8}$ |
| | $A_{FB}$ | | |
| | 1st Scenario | 2nd Scenario | 3rd Scenario |
| In $q^2 = 6$ | $(0.009 \pm 0.004)$ | $(0.021 \pm 0.007)$ | $(0.010 \pm 0.004)$ |
| In $q^2 = 12$ | $(0.023 \pm 0.010)$ | $(0.033 \pm 0.026)$ | $(0.032 \pm 0.012)$ |
| In $q^2 = 18$ | $(0.129 \pm 0.016)$ | $(0.228 \pm 0.032)$ | $(0.152 \pm 0.019)$ |

## VI. Conclusion:

Universality of electroweak couplings for the 3 lepton families is successfully explained by SM theory. Although the signals of LFV have not been perfectly recognized in LHCb, yet this advanced analysis could evidently be mesmerizing in the horizon of high energy physics that might fabricate the future of BSM physics more vibrantly. The transition $b \to s l_i^+ l_j^-$ is previously studied in leptoquark model and MSSM. In this paper, we have studied this quark transition for lepton flavour violating decays of $\Lambda_b$ baryon. To the best of our knowledge these LFV decays have not studied experimentally till now but some NP models have discussed [18-30]. Motivated by the results of references [23, 24], we have studied $\Lambda_b \to \Lambda l_i^+ l_j^-$ decays in the context of non-universal $Z'$ model. While analysing we have constrained the flavour changing coupling $B_{sb}$ from $B_s - \bar{B}_s$ mixing and LFV couplings from various LFV decays. The constraint on $\mu - e - Z'$ coupling is the order of $10^{-3}$ whereas $\tau - \mu - Z'$ coupling is the order of $10^{-1}$ and $B_{sb}$ is fixed for three scenarios [shown at Table 1]. In the 2nd scenario the predicted values of the observables are higher due to enhancement of NP contribution. The zero crossings are present in $A_{FB}$ for both decays but it is more prominent for $\Lambda_b \to \Lambda \tau^+ \mu^-$ decay.

The observables, which are discussed in this work, are also explained with the effects of vector and scalar leptoquarks in the references [24] and [23] respectively. Two observables: differential branching fractions and forward backward asymmetries are studied in those works where ref. [24] found the maximum values of branching ratio and lepton side forward backward asymmetry for $\Lambda_b \to \Lambda \tau^+ \mu^-$ decay over the whole $q^2$ region as $(7.83 \times 10^{-6})$ and $(-0.2504)$ respectively whereas scalar leptoquark model [23] found the branching fraction for $\Lambda_b \to \Lambda \tau^- \mu^+$ decay of the order of $10^{-10} - 10^{-9}$. We have found the differential branching ratio for $\Lambda_b \to \Lambda \tau^+ \mu^-$ decay in non-universal $Z'$ model over the whole kinematic region as ($2.46 \times 10^{-7}$) for scenario 1, ($7.91 \times 10^{-7}$) for scenario 2 and ($3.39 \times 10^{-7}$) for scenario 3. Another powerful observable to look for NP is the zero crossing position of lepton side forward backward asymmetry. The shifting of this position is very sensitive to the physics beyond the SM. Here, we have found that the zero crossing is shifting with different NP scenarios in Fig. 3d and Fig.6d in a magnified way and this infers the responsiveness of NP on the observable. Vector leptoquark model has obtained the zero crossing in between $q^2 = 8$ GeV$^2$ to $q^2 = 9$ GeV$^2$ whereas we have obtained at $q^2 = 6.5$ GeV$^2$ for scenario 1, $q^2 = 4.0$ GeV$^2$ for scenario 2 and $q^2 = 6.2$ GeV$^2$ for scenario 3 for $\Lambda_b \to \Lambda \tau^+ \mu^-$ decay. It is noted that the higher value of NP coupling shifts the zero crossing nearer to origin.

According to our calculation the larger values of the observables are obtained at high $q^2$ regime with magnified contribution of $Z'$ boson. The bands of the figures 2, 3, 5, 6 interpret that the uncertainty at high $q^2$ region is much lower than the low $q^2$ region. It can be expected that the predicted values of $\frac{d\mathcal{B}}{dq^2}$ and $A_{FB}$ at Table 2 and Table 3 would help experimental community to access it in the near future.

**Acknowledgement:** We thank the reviewer for suggesting valuable improvements of our manuscript. S. Biswas acknowledges NIT Durgapur for providing fellowship for her research work. P. Nayek and S. Sahoo would like to thank SERB, DST, Govt. of India for financial support through grant no. EMR/2015/000817. P. Maji thanks DST, Government of India for providing INSPIRE Fellowship during her research. We would like to thank Dr. Sunando Patra, Assistant Professor, Bangabasi Evening College, Kolkata, India, for useful discussions.

**Appendix A:**

Being comparable to the effective Hamiltonian the expressions of transversity amplitudes becomes as,

$$A_{\perp 1}^{L,(R)} = -\sqrt{2} N \left( f_\perp^V \sqrt{2s_-} (C_9' \mp C_{10}') \right), \tag{A1}$$

$$A_{\parallel 1}^{L,(R)} = \sqrt{2} N \left( f_\perp^A \sqrt{2s_+} (C_9' \mp C_{10}') \right), \tag{A2}$$

$$A_{\perp 0}^{L,(R)} = \sqrt{2} N \left( f_0^V (m_{\Lambda_b} + m_\Lambda) \sqrt{\frac{s_-}{q^2}} (C_9' \mp C_{10}') \right), \tag{A3}$$

$$A_{\parallel 0}^{L,(R)} = -\sqrt{2} N \left( f_0^A (m_{\Lambda_b} - m_\Lambda) \sqrt{\frac{s_+}{q^2}} (C_9' \mp C_{10}') \right), \tag{A4}$$

$$A_{\perp t} = -2\sqrt{2} N f_t^V (m_{\Lambda_b} - m_\Lambda) \sqrt{\frac{s_+}{q^2}} C_{10}', \tag{A5}$$

$$A_{\parallel t} = 2\sqrt{2} N f_t^A (m_{\Lambda_b} + m_\Lambda) \sqrt{\frac{s_-}{q^2}} C_{10}', \tag{A6}$$

here, $N = G_F \alpha V_{tb} V_{ts}^* \sqrt{\tau_{\Lambda_b} q^2 \frac{\sqrt{\lambda(m_{\Lambda_b}^2, m_\Lambda^2, q^2)}}{2^{15} m_{\Lambda_b}^3 \pi^5}} \beta\beta'$ .

In general $\lambda(a,b,c) = a^2 + b^2 + c^2 - 2(ab + bc + ac)$ is the triangular function and the other expressions are given as below,

$$\begin{aligned} s_+ &= \{(m_{\Lambda_b} + m_\Lambda)^2 - q^2\} \\ s_- &= \{(m_{\Lambda_b} - m_\Lambda)^2 - q^2\} \end{aligned} \quad . \tag{A7}$$

**Appendix B:**

To parametrize the $\Lambda_b \to \Lambda$ hadronic matrix elements we have chosen the helicity basis [49, 51] in terms of the matrix elements for the vector and axial-vector current. The matrix element for vector current is,

$$\begin{aligned} &\langle \Lambda(k, s_k) | \bar{s}\gamma^\mu b | \Lambda(p, s_p) \rangle \\ &= \bar{u}(k, s_k) \Bigg[ f_t^V(q^2)(m_{\Lambda_b} - m_\Lambda) \frac{q^\mu}{q^2} \\ &\quad + f_0^V(q^2) \frac{(m_{\Lambda_b} + m_\Lambda)}{s_+} \left\{ p^\mu + k^\mu - \frac{q^\mu}{q^2}(m_{\Lambda_b}^2 - m_\Lambda^2) \right\} \\ &\quad + f_\perp^V(q^2) \left\{ \gamma^\mu - \frac{2m_\Lambda}{s_+} p^\mu - \frac{2m_{\Lambda_b}}{s_+} k^\mu \right\} \Bigg] u(p, s_p) \, , \end{aligned} \tag{B1}$$

and matrix element for axial-current is,

$$\begin{aligned} &\langle \Lambda(k, s_k) | \bar{s}\gamma^\mu \gamma_5 b | \Lambda(p, s_p) \rangle \\ &= -\bar{u}(k, s_k)\gamma_5 \Bigg[ f_t^A(q^2)(m_{\Lambda_b} + m_\Lambda) \frac{q^\mu}{q^2} \\ &\quad + f_0^A(q^2) \frac{(m_{\Lambda_b} - m_\Lambda)}{s_-} \left\{ p^\mu + k^\mu - \frac{q^\mu}{q^2}(m_{\Lambda_b}^2 - m_\Lambda^2) \right\} \\ &\quad + f_\perp^A(q^2) \left\{ \gamma^\mu + \frac{2m_\Lambda}{s_-} p^\mu - \frac{2m_{\Lambda_b}}{s_-} k^\mu \right\} \Bigg] u(p, s_p) \, . \end{aligned} \tag{B2}$$

The $q^2$ dependence fit function for which we have used a higher order fit is,

$$f(q^2) = \frac{1}{1 - q^2/\left(m_{pole}^f\right)^2} \left[ a_0^f + a_1^f z(q^2, t_+) + a_2^f \left(z(q^2, t_+)\right)^2 \right] \, , \tag{B3}$$

where, $z(q^2, t_+) = \frac{\sqrt{t_+ - q^2} - \sqrt{t_+ - t_0}}{\sqrt{t_+ - q^2} + \sqrt{t_+ - t_0}}$ , $t_+ = (m_B + m_K)^2$ and $t_0 = (m_{\Lambda_b} - m_\Lambda)^2$ .

The numerical values of the fit parameters are recorded in Table B1 and B2 [51]. In our analysis we have taken the central values of the fit parameters.

## Table B1: Values of B meson pole masses $m^f_{pole}$

| $f$ | $m^f_{pole}$ [GeV] |
|---|---|
| $f^V_0, f^V_\perp$ | 5.416 |
| $f^V_t$ | 5.711 |
| $f^A_0, f^A_\perp$ | 5.750 |
| $f^A_t$ | 5.367 |

## Table B2: Central values of the higher order form factor parameters

| $f$ | $a_0$ | $a_1$ | $a_2$ |
|---|---|---|---|
| $f^V_0$ | 0.4229 | −1.3728 | 1.7972 |
| $f^V_t$ | 0.3604 | −0.9248 | 0.9861 |
| $f^V_\perp$ | 0.5148 | −1.4781 | 1.2496 |
| $f^A_0$ | 0.3522 | −1.2968 | 2.7106 |
| $f^A_t$ | 0.4059 | −1.1622 | 1.1490 |
| $f^A_\perp$ | 0.3522 | −1.3607 | 2.4621 |

**Appendix C:**

The leptonic helicity amplitudes can be written explicitly as,

$$\bar{\epsilon}^\mu(\lambda)\bar{u}_{l_j}\gamma_\mu(1 \mp \gamma_5)v_{l_i}, \tag{C1}$$

From the ref [84] the explicit expressions of the spinor for lepton $l_j^-$ is expressed as,

$$\bar{u}_{l_j}(\lambda) = \begin{bmatrix} \sqrt{E_l + m_l}\chi^u_\lambda \\ 2\lambda\sqrt{E_l - m_l}\chi^u_\lambda \end{bmatrix}, \chi^u_{+\frac{1}{2}} = \begin{bmatrix} \cos\frac{\theta_l}{2} \\ \sin\frac{\theta_l}{2} \end{bmatrix}, \chi^u_{-\frac{1}{2}} = \begin{bmatrix} -\sin\frac{\theta_l}{2} \\ \cos\frac{\theta_l}{2} \end{bmatrix}. \tag{C2}$$

Another spinor for the lepton $l_i^+$ which is moving opposite to lepton $l_j^-$,

$$\bar{v}_{l_i}(\lambda) = \begin{bmatrix} \sqrt{E_l - m_l}\chi^v_{-\lambda} \\ -2\lambda\sqrt{E_l + m_l}\chi^v_{-\lambda} \end{bmatrix}, \chi^v_{+\frac{1}{2}} = \begin{bmatrix} \sin\frac{\theta_l}{2} \\ -\cos\frac{\theta_l}{2} \end{bmatrix}, \chi^v_{-\frac{1}{2}} = \begin{bmatrix} \cos\frac{\theta_l}{2} \\ \sin\frac{\theta_l}{2} \end{bmatrix}. \tag{C3}$$

From the ref [84] we have studied the two component spinors which are related as $\chi^v_{-\lambda} = \xi_\lambda \chi^u_\lambda$ and $\xi_\lambda = 2\lambda e^{-2i\lambda\varphi}$ where $\varphi$ is the azimuthal angle.

Following all these considerations we have obtained the expressions of the lepton helicity amplitudes $L_{L(R),\lambda}^{\lambda_j,\lambda_i}$ which are collected below,

$$L_{L,+1}^{+\frac{1}{2}+\frac{1}{2}} = \frac{1}{\sqrt{2}}\left[m_i(\beta'+\beta) + m_j(\beta'-\beta)\right]\sin\theta_l \ , \tag{C4}$$

$$L_{L,+1}^{+\frac{1}{2}-\frac{1}{2}} = -\sqrt{\frac{q^2}{2}}(\beta'-\beta)(1-\cos\theta_l) \ , \tag{C5}$$

$$L_{L,+1}^{-\frac{1}{2}+\frac{1}{2}} = \sqrt{\frac{q^2}{2}}(\beta'-\beta)(1+\cos\theta_l) \ , \tag{C6}$$

$$L_{L,+1}^{-\frac{1}{2}-\frac{1}{2}} = -\frac{1}{\sqrt{2}}\left[m_i(\beta'-\beta) + m_j(\beta'+\beta)\right] \ , \tag{C7}$$

$$L_{R,+1}^{+\frac{1}{2}+\frac{1}{2}} = \frac{1}{\sqrt{2}}\left[m_i(\beta'-\beta) + m_j(\beta'+\beta)\right]\sin\theta_l \ , \tag{C8}$$

$$L_{R,+1}^{+\frac{1}{2}-\frac{1}{2}} = -\sqrt{\frac{q^2}{2}}(\beta'-\beta)(1-\cos\theta_l) \ , \tag{C9}$$

$$L_{R,+1}^{-\frac{1}{2}+\frac{1}{2}} = \sqrt{\frac{q^2}{2}}(\beta'-\beta)(1+\cos\theta_l) \ , \tag{C10}$$

$$L_{R,+1}^{-\frac{1}{2}-\frac{1}{2}} = -\frac{1}{\sqrt{2}}\left[m_i(\beta'+\beta) + m_j(\beta'-\beta)\right]\sin\theta_l \ , \tag{C11}$$

$$L_{L,-1}^{+\frac{1}{2}+\frac{1}{2}} = -\frac{1}{\sqrt{2}}\left[m_i(\beta'+\beta) + m_j(\beta'-\beta)\right] \ , \tag{C12}$$

$$L_{L,-1}^{+\frac{1}{2}-\frac{1}{2}} = -\sqrt{\frac{q^2}{2}}(\beta'-\beta)(1+\cos\theta_l) \ , \tag{C13}$$

$$L_{L,-1}^{-\frac{1}{2}+\frac{1}{2}} = \sqrt{\frac{q^2}{2}}(\beta'+\beta)(1-\cos\theta_l) \ , \tag{C14}$$

$$L_{L,-1}^{-\frac{1}{2}-\frac{1}{2}} = \frac{1}{\sqrt{2}}\left[m_i(\beta'-\beta) + m_j(\beta'+\beta)\right]\sin\theta_l \ , \tag{C15}$$

$$L_{R,-1}^{+\frac{1}{2}+\frac{1}{2}} = -\frac{1}{\sqrt{2}}\left[m_i(\beta'-\beta) + m_j(\beta'+\beta)\right]\sin\theta_l \ , \tag{C16}$$

$$L_{R,-1}^{+\frac{1}{2}-\frac{1}{2}} = \sqrt{\frac{q^2}{2}}(\beta'+\beta)(1+\cos\theta_l) \ , \tag{C17}$$

$$L_{R,-1}^{-\frac{1}{2}+\frac{1}{2}} = \sqrt{\frac{q^2}{2}}(\beta'-\beta)(1-\cos\theta_l) \ , \tag{C18}$$

$$L_{R,-1}^{-\frac{1}{2}-\frac{1}{2}} = \frac{1}{\sqrt{2}}\left[m_i(\beta'+\beta) + m_j(\beta'-\beta)\right]\sin\theta_l \ , \tag{C19}$$

$$L_{L,0}^{+\frac{1}{2}+\frac{1}{2}} = -[m_i(\beta' + \beta) - m_j(\beta' - \beta)]\cos\theta_l ,\tag{C20}$$

$$L_{L,0}^{+\frac{1}{2}-\frac{1}{2}} = \sqrt{q^2}(\beta' - \beta)\cos\theta_l ,\tag{C21}$$

$$L_{L,0}^{-\frac{1}{2}+\frac{1}{2}} = \sqrt{q^2}(\beta' + \beta)\cos\theta_l ,\tag{C22}$$

$$L_{L,0}^{-\frac{1}{2}-\frac{1}{2}} = [m_i(\beta' - \beta) + m_j(\beta' + \beta)]\cos\theta_l ,\tag{C23}$$

$$L_{R,0}^{+\frac{1}{2}+\frac{1}{2}} = -[m_i(\beta' - \beta) + m_j(\beta' + \beta)]\cos\theta_l ,\tag{C24}$$

$$L_{R,0}^{+\frac{1}{2}-\frac{1}{2}} = \sqrt{q^2}(\beta' + \beta)\cos\theta_l ,\tag{C25}$$

$$L_{R,0}^{-\frac{1}{2}+\frac{1}{2}} = \sqrt{q^2}(\beta' - \beta)\cos\theta_l ,\tag{C26}$$

$$L_{R,0}^{-\frac{1}{2}-\frac{1}{2}} = [m_i(\beta' + \beta) + m_j(\beta' - \beta)]\cos\theta_l ,\tag{C27}$$

$$L_{L,t}^{+\frac{1}{2}+\frac{1}{2}} = [m_i(\beta' + \beta) + m_j(\beta' - \beta)] ,\tag{C28}$$

$$L_{L,t}^{+\frac{1}{2}-\frac{1}{2}} = L_{L,t}^{-\frac{1}{2}+\frac{1}{2}} = 0 ,\tag{C29}$$

$$L_{L,t}^{-\frac{1}{2}-\frac{1}{2}} = [m_i(\beta' - \beta) + m_j(\beta' + \beta)] ,\tag{C30}$$

$$L_{R,t}^{+\frac{1}{2}+\frac{1}{2}} = -[m_i(\beta' - \beta) + m_j(\beta' + \beta)] ,\tag{C31}$$

$$L_{R,t}^{+\frac{1}{2}-\frac{1}{2}} = L_{R,t}^{-\frac{1}{2}+\frac{1}{2}} = 0 ,\tag{C32}$$

$$L_{R,t}^{-\frac{1}{2}-\frac{1}{2}} = -[m_i(\beta' + \beta) + m_j(\beta' - \beta)] ,\tag{C33}$$

**Appendix D:**

Input values which are used in the investigation are recorded in the following table.

| Parameter | Values |
|---|---|
| $m_\mu$ | $(105.66 \pm 0.0000024)$ MeV |
| $m_e$ | $(0.51 \pm 0.0000000031)$ MeV |
| $m_\tau$ | $(1776.86 \pm 0.12)$ MeV |
| $m_B$ | $(5279.55 \pm 0.26)$ MeV |
| $m_K$ | $(497.611 \pm 0.013)$ MeV |
| $m_\Lambda$ | $(1115.683 \pm 0.006)$ MeV |
| $m_{\Lambda_b}$ | $(5619.60 \pm 0.17)$ MeV |
| $G_F$ | $(1.166 \pm 0.0000006) \times 10^{-5}$ GeV$^{-2}$ |
| $\|V_{tb}\|$ | $(1.019 \pm 0.025)$ |
| $\|V_{ts}\|$ | $(39.4 \pm 2.3) \times 10^{-3}$ |